\documentclass[12pt]{iopart}
\usepackage{iopams}
\usepackage{graphicx}
\usepackage{color}
\usepackage{verbatim}
\begin{document}
\title{Relaxation at finite temperature in Fully-Frustrated Ising Models}
\author{J.-C. Walter and C. Chatelain}
\address{
Institute for Theoretical Physics,\\
KULeuven, Celestijnenlaan 200D,\\
B-3001 Leuven, Belgium}
\address{
Groupe de Physique Statistique,\\
Institut Jean Lamour, UMR 7198,\\
Nancy-Universit\'e, CNRS,
BP~70239, Boulevard des aiguillettes,\\
F-54506 Vand{\oe}uvre l\`es Nancy Cedex, France}
\ead{chatelai@ijl.nancy-universite.fr}

\pacs{05.10.Ln, 05.50.+q, 05.70.Jk, 05.70.Ln}

\begin{abstract}
We consider by means of Monte Carlo simulations the relaxation in the
paramagnetic phase of the anti-ferromagnetic Ising model on a
triangular lattice and of a fully-frustrated Ising model on a square
lattice. In contradistinction to previous studies of the second
model, we show that spin-spin correlation functions do not decay with
a stretched-exponential law at low temperature but that both models
display an exponential decay with logarithmic corrections that are
interpreted as the signature of topological defects.
\end{abstract}
\date{\today}
\maketitle

\section{Introduction}
The study of relaxation in frustrated systems is of particular
interest because their dynamics may become anomalously slow at low
temperature and eventually freeze below a certain
temperature. The paradigmatic example of such a slow dynamics is
given by 
spin glasses which combine both frustration and randomness.
Frustration alone does not necessarily imply a glassy
dynamics. Even though each one of its plaquettes is frustrated, the
anti-ferromagnetic Ising model on a triangular
lattice~\cite{PhysRevB.7.5017} (AFIM) was shown to display the same
dynamics as unfrustrated systems~\cite{zeng_zero-temperature_1997}.
Interestingly, if elastic deformations of the lattice are allowed
and coupled to the spin degrees of freedom, the relaxation is well
described by a (stretched-exponential) Kohlrausch-Williams-Watts law
like in glasses~\cite{chen_elastic_1986,gu_monte_1996,yin_slow_2002}.
This behavior is observed experimentally for example with closely-packed
colloidal spheres~\cite{han_geometric_2008}.\\

In this paper, we consider both the AFIM and the Fully-Frustrated
Ising Model (FFIM) defined on a square lattice by the Hamiltonian
     \begin{equation}
       -\beta H=\sum_{x,y} \big[\sigma_{x,y}\sigma_{x+1,y}
       +(-1)^{f(x,y)}\sigma_{x,y}\sigma_{x,y+1}\big],\quad\quad\sigma_{x,y}=\pm 1
     \end{equation}
where $f(x,y)=x+y$ in the so-called zigzag model and $f(x,y)=x$ in
pileup-domino configuration. In both cases, each plaquette of the
square lattice contains an odd number of anti-ferromagnetic bonds and,
as a consequence, is frustrated. The pileup-domino configuration
allows for exact diagonalization of the transfer matrix by free
fermion techniques~\cite{villain_spin_1977}. Both AFIM and FFIM are
known to belong to the same universality class. At the critical
temperature $T_c=0$, spin-spin autocorrelation functions decay
algebraically
        \begin{equation}
        C(t,s)=\langle\sigma(t)\sigma(s)\rangle
        \sim (t-s)^{-\eta/z}
        \label{ScalCritic}
        \end{equation}
where $\eta=1/2$ and $z=2$ is the dynamical
exponent. In the case of the AFIM, evidences have been given of the existence
of topological defects interacting through a logarithmic Coulombian
potential in the paramagnetic phase~\cite{moore_vortex_1999,yin_effective_2000}.
While no signature of these defects was observed
in earlier simulations of the AFIM~\cite{kim_nonequilibrium_2003},
we have shown that such defects manifest themselves in the
FFIM~\cite{walter_logarithmic_2008}.
When the system is initially prepared in the paramagnetic phase and
then quenched at $T_c=0$, they pin the domain walls and slow
down their motion so that aging takes place in the same way as in
homogeneous systems but with logarithmic corrections. The correlation
length grows as $\xi(t)\sim (t/\ln t)^{1/z}$ and spin-spin
autocorrelation functions behave as
       \begin{equation}
       C(t,s)\sim s^{-\eta/z}\left({t\ln s\over s\ln t}
       \right)^{-\lambda/z}
       \label{ScalAging}
       \end{equation}
where both $z$ and $\lambda$ are compatible with $2$.
\\

This result suggests that at finite temperature, the relaxation
should be exponential, like in unfrustrated systems, but with
logarithmic corrections due to the presence of topological
defects. However, on the basis of Monte Carlo simulations,
a stretched-exponential relaxation~\footnote{The exponent in the
stretched-exponential law is usually denoted $\beta$. We will use the
notation $\kappa$ instead in order to avoid confusion with the
inverse temperature $\beta=1/k_BT$.}
      \begin{equation}
        C(t)\sim e^{-(t/\tau)^\kappa}
        \label{eqStr}
      \end{equation}
with a temperature-dependent exponent $\kappa$ has been reported in the
FFIM below the temperature $T_p\simeq 1.701$ at which Kasteleyn-Fortuin clusters
start to percolate~\cite{fierro_percolation_1999,franzese_precursor_1999}.
In this work, we reconsider the AFIM and the FFIM and show that
Monte Carlo data are in better agreement with an exponential decay
when taking into account logarithmic corrections than with a stretched
exponential. In the first section, a dynamical scaling hypothesis is
set up to predict the behavior of equilibrium spin-spin autocorrelation
functions in fully-frustrated Ising models. The expression is then compared
with Monte Carlo data for the AFIM in the second section and the FFIM in
the third one.

\section{Dynamical scaling of spin-spin autocorrelation functions}
We analyze the two-time spin-spin correlation functions in the
framework of dynamical scaling~\cite{hohenberg_theory_1977}.
Upon a dilatation with a scale factor $b$, the equilibrium correlation
$C(\vec r,t,T)=\langle\sigma_0(0)\sigma_{\vec r}(t)\rangle$ at temperature $T$
is assumed to satisfy the homogeneity relation
      \begin{equation}
        C(\vec r,1/t,|T-T_c|)=b^{-2x_\sigma}C\big(r/b,
        b^z/t,|T-T_c|b^{1/\nu}\big)
        \label{eq1}
        \end{equation}
where $x_\sigma$ is the scaling dimension of magnetization density with
$2x_\sigma=\eta$ for two-dimensional systems and $z$ is the dynamical
exponent. The motivation for the last two arguments of the scaling function
in equation (\ref{eq1}) comes for the behavior of the correlation length
either with time, $\xi\sim t^{1/z}$, or with temperature,
$\xi\sim |T-T_c|^{-1/\nu}$. Letting $b=t^{1/z}$ in equation (\ref{eq1}),
we obtain
       \begin{equation}
         C(\vec r,t)=t^{-\eta/z}{\cal C}\big(r/t^{1/z},|T-T_c|t^{1/\nu
           z}\big)\label{ScalingHyp}
       \end{equation}
The algebraic prefactor corresponds to the critical behavior while
the scaling function includes all corrections to it.
The characteristic time
        \begin{equation}
          \tau\sim \xi^z\sim|T-T_c|^{-\nu z}
          \label{ScalTau}
          \end{equation}
appears as the relaxation time of the system. In the following, we are
interested only in autocorrelation functions, i.e. $r=0$. Moreover, we
expect an exponential decay of the scaling function ${\cal C}(t/\tau)$
in the paramagnetic phase. Therefore, the autocorrelation function generally
reads at equilibrium
       \begin{equation}
         C(t,T)\sim {e^{-t/\tau}\over t^{\eta/z}}
         \label{eq2}
       \end{equation}

Two modifications need to be made to apply this hypothesis to the
AFIM and the FFIM. First, as already mentioned, the existence of
topological defects slows down the motion of domain walls and thus
the growth of the correlation length, i.e. $\xi\sim (t/\ln t)^{1/z}$.
A logarithmic correction has to be included in the scaling hypothesis
(\ref{ScalingHyp})
       \begin{equation}
         C(0,t)=t^{-\eta/z}{\cal C}\big(0,|T-T_c|(t/\ln t)^{1/\nu
           z}\big)\sim {e^{-t/\tau\ln t}\over t^{\eta/z}}
         \label{eq4}
       \end{equation}
The algebraic decay in front of the scaling function is not affected by
logarithmic corrections because it describes the critical behavior
for which topological defects are paired. In contradistinction, 
the scaling function ${\cal C}$ corresponds to the deviation to this
behavior caused by non-vanishing scaling fields. The latter brings the
system out-of-criticality, i.e. into the paramagnetic phase where free
topological defects are encountered. ${\cal C}$ should thus involve
logarithmic corrections. Note that the same kind of behavior was
assumed during aging (\ref{ScalAging}). The second modification
concerns the relaxation time. Since the correlation length is
known~\cite{forgacs_ground-state_1980,lukic_finite-size_2006}
to diverge exponentially with
temperature, i.e. $\xi\sim e^{2/T}$ and not algebraically, equation
(\ref{ScalTau}) has to be replaced by
        \begin{equation}
          \tau\sim \xi^z\sim e^{2z/T}.
          \label{eq5}
          \end{equation}

\section{Relaxation of the AFIM}
We have studied the AFIM and FFIM by means of large-scale Monte Carlo
simulations for a two-dimensional lattice with $192\times 192$ sites. The
dynamics is the heat-bath local Markovian process introduced by
Glauber~\cite{glauber_time-dependent_1963}. We shall consider first
the AFIM. Inverse temperatures $\beta=1/k_BT$ in the range $[0.75;3.00]$ have
been considered. Data have been averaged over $30,000$ independent
histories. Error bars on the average correlation $C(t,s)$ have been
estimated at each times $t$ and $s$ as the standard deviation of the
data produced during the different independent histories.\\

\begin{center}
\begin{figure}
\centerline{\includegraphics[width=12cm]{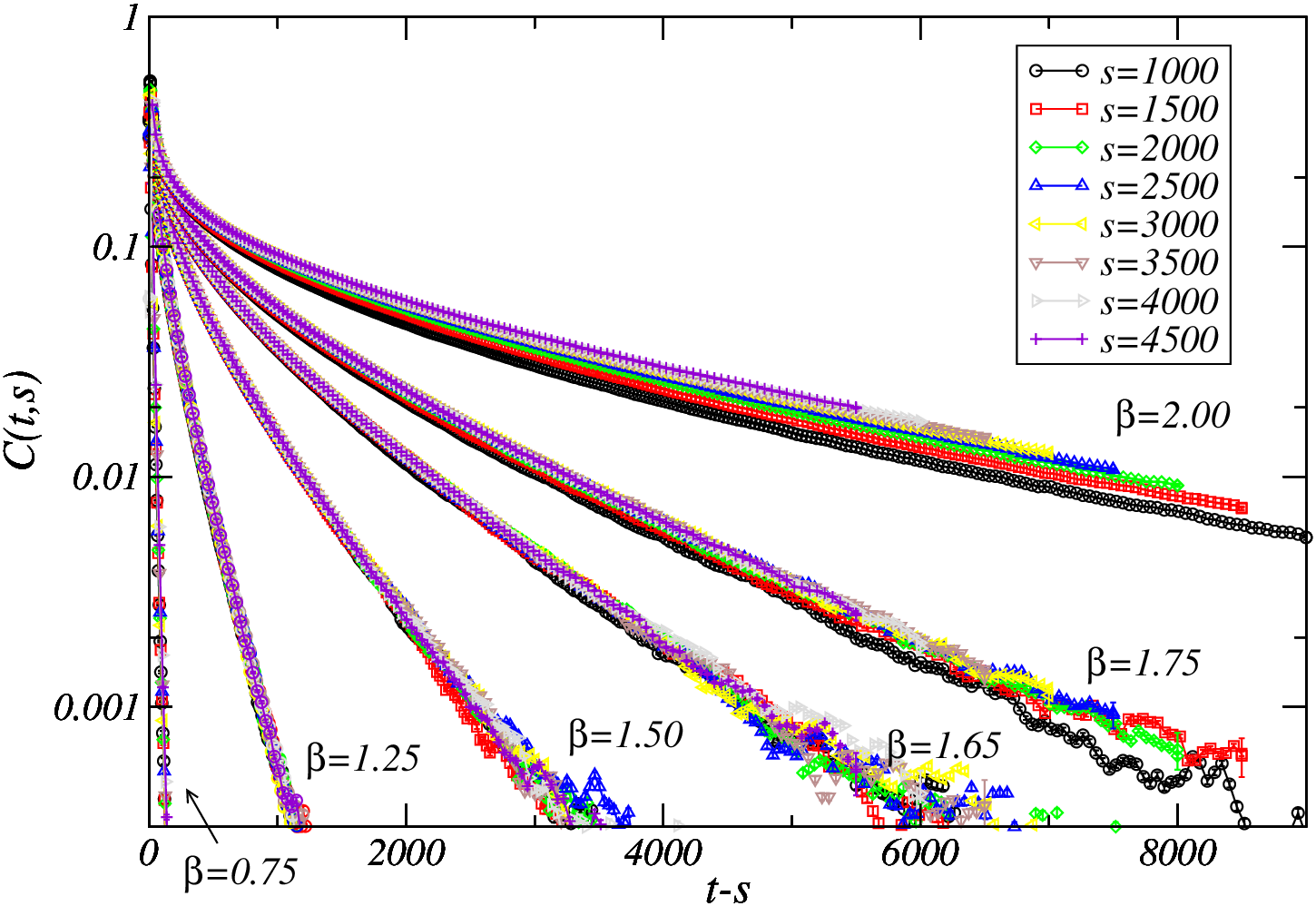}}
\caption{Relaxation of the spin-spin correlation function $C(t,s)$
of the AFIM versus $t-s$ at six different inverse temperatures
$\beta=0.75$, $1.25$, $1.50$, $1.65$, $1.75$, and $2.00$.
The different curves correspond to different times $s$.
Error bars are represented but they are hardly visible on the figure
because they are of order ${\cal O}(10^{-5})$, i.e. much smaller
than the symbols.}
\label{Fig1}
\end{figure}
\end{center}

To check that the system had thermalized when the
measurements were started, we monitored the two-time spin-sin
correlation functions $C(t,s)$ which are expected to depend only on
$t-s$ in the stationary state. On figure \ref{Fig1}, spin-spin
autocorrelation functions of the AFIM are plotted at different
temperatures with respect to $t-s$. For sufficiently large values of $s$,
one can observe the collapse of the curves, indicating that the system
has reached equilibrium. In the example of the data presented in
figure \ref{Fig1}, one is led to the conclusion that below the inverse
temperature $\beta=1/k_BT=1.75$, equilibrium has already been reached
at time $s=1000$. At $\beta=1.75$, equilibrium is reached only at time
$s=2000$ while for larger values of $\beta$, no collapse is observed
indicating that equilibrium is not reached yet at time $s=4500$.
For this reason, only inverse temperatures $\beta\le 1.75$ with
$s\ge 2000$ will be considered in the following.

\begin{center}
\begin{figure}
\centerline{\includegraphics[width=12cm]{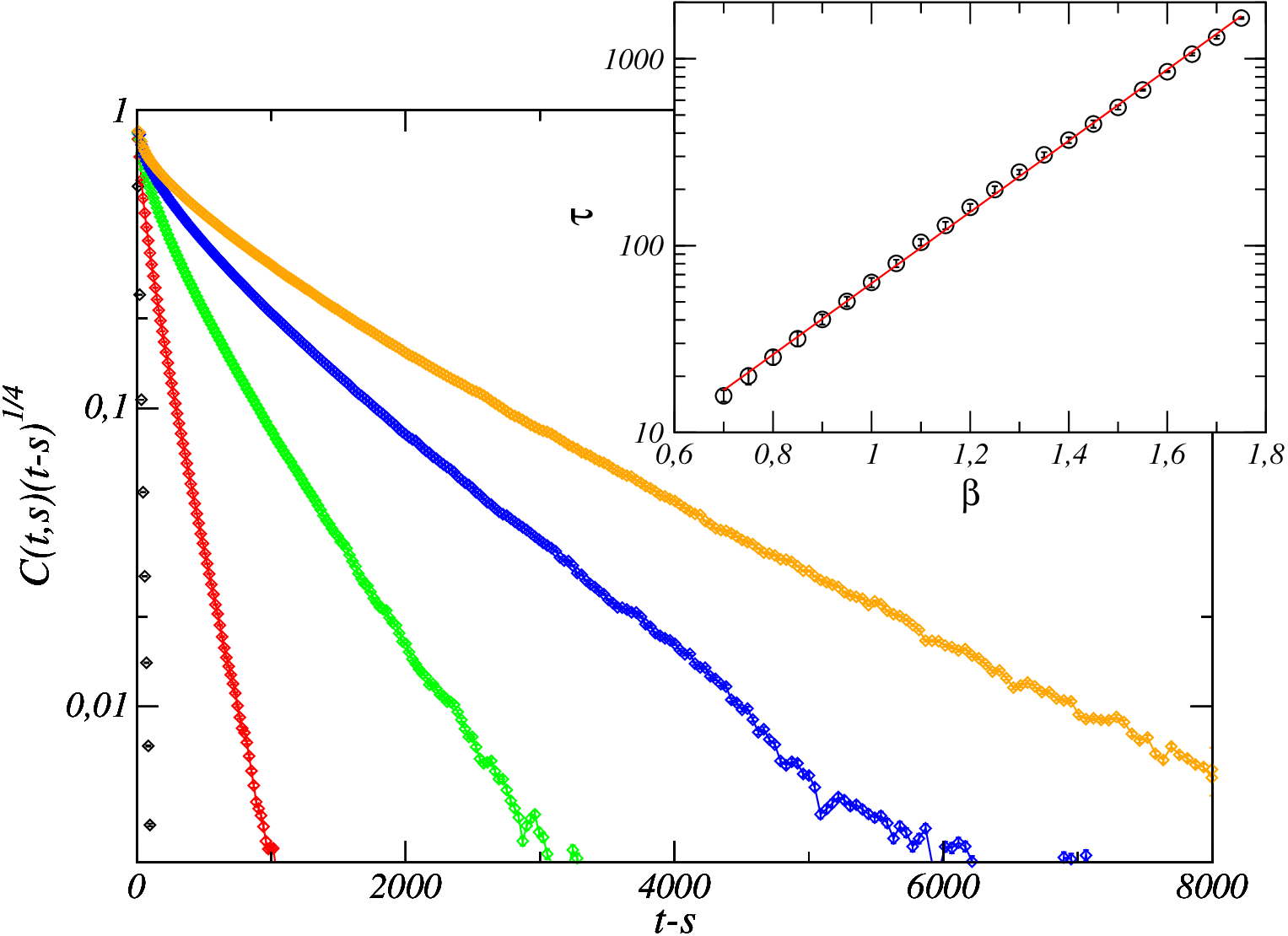}}
\caption{Scaling function $C(t,s)(t-s)^{1/4}$ at time $s=2000$ versus
  $t-s$ for the AFIM at different inverse temperatures $\beta=0.75$, $1.25$,
  $1.50$, $1.65$, and $1.75$ (from left to right). In the inset, the
  relaxation time $\tau$ obtained by interpolation of the scaling
  function as $e^{-(t-s)/\tau}$ is plotted versus the inverse
  temperature $\beta$. The straight line is the interpolated
  behavior $\tau\sim e^{4.31(2)\beta}$.}
\label{Fig2}
\end{figure}
\end{center}

We have tested three possible scenarii: exponential decay of the
spin-spin autocorrelation functions (\ref{eq2}), exponential decay
with logarithmic corrections (\ref{eq4}), and stretched exponential
(\ref{eqStr}). On figure \ref{Fig2},
the scaling function $C(t,s)(t-s)^{1/4}$ is plotted versus $t-s$ with
a waiting time $s=2000$ (ensuring equilibration as discussed
above). In the scenario (\ref{eq2}), this function is expected to
decay exponentially. This behavior is indeed observed over a large
range of times $t-s$ for all inverse temperatures $\beta\le 1.75$. 
We estimated the relaxation time by interpolation of the scaling function
as $e^{-t/\tau}$ with a sliding interpolation window. As can be
seen in the inset of figure \ref{Fig2}, the dependence of $\tau$ on $\beta$
is well described by an exponential, as expected for the AFIM (see
equation \ref{eq5}). A fit gives $\tau\sim e^{4.31(2)\beta}$, a
behavior which is close to the expected one (\ref{eq5}), thought
significantly outside error bars.

\begin{center}
\begin{figure}
\centerline{\includegraphics[width=12cm]{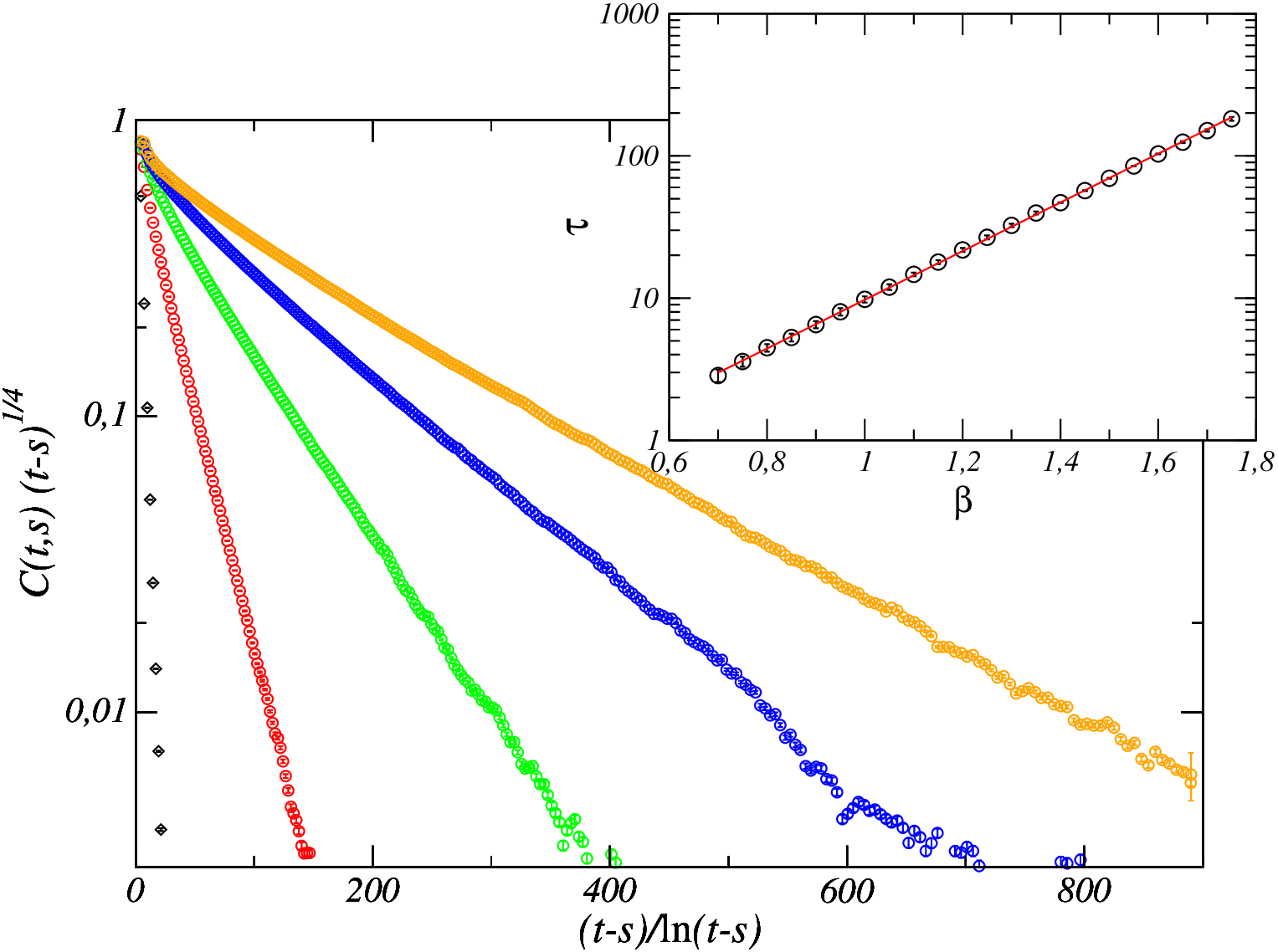}}
\caption{Scaling function $C(t,s)(t-s)^{1/4}$ at time $s=2000$ versus
  $(t-s)/\ln(t-s)$ for the AFIM at different inverse temperatures
  $\beta=0.75$, $1.25$, $1.50$, $1.65$, and $1.75$
  (from from left to right). In the inset, the
  relaxation time $\tau$ obtained by interpolation of the scaling
  function as $e^{-(t-s)/\tau\ln(t-s)}$ is plotted versus the inverse
  temperature $\beta$. The straight line is the interpolated
  behavior $\tau\sim e^{3.92(2)\beta}$.}
\label{Fig3}
\end{figure}
\end{center}

On figure \ref{Fig3}, the second scenario (\ref{eq4}) is tested. The
scaling function $C(t,s)(t-s)^{1/4}$ is plotted versus
$(t-s)/\ln(t-s)$. The interpolation with an exponential gives the
values represented in the inset. As before, the relaxation time behaves
exponentially with the inverse temperature. A fit gives the law
$\tau\sim e^{3.92(2)\beta}$, a behavior which is much closer to
(\ref{eq5}) than without logarithmic corrections, even though still
outside error bars.

\begin{center}
\begin{figure}
\centerline{\includegraphics[width=12cm]{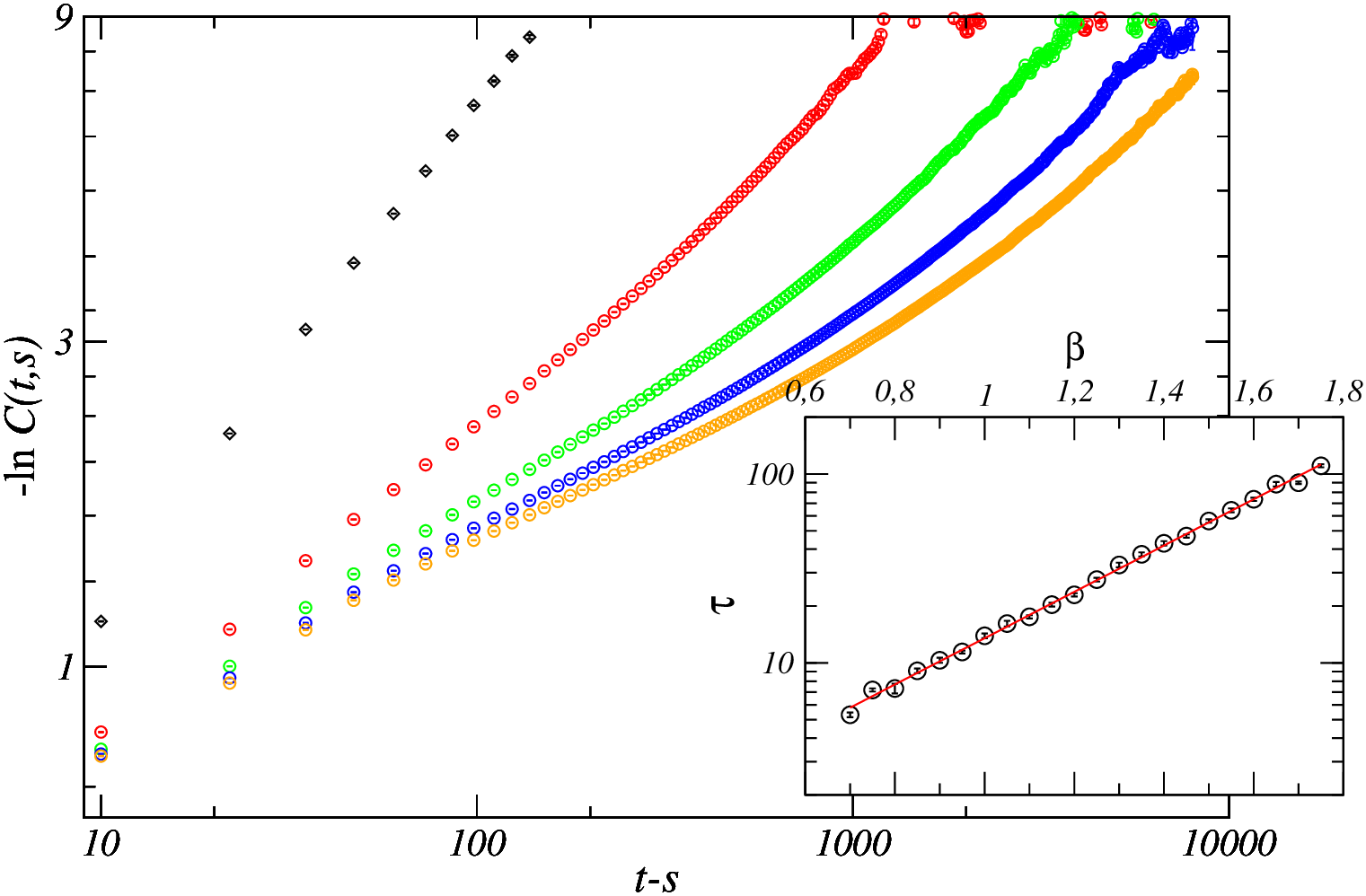}}
\caption{Scaling function $-\ln C(t,s)$ at time $s=2000$ versus $t-s$
  for the AFIM at different inverse temperatures $\beta=0.75$, $1.25$,
  $1.50$, $1.65$ and $1.75$ (from top to bottom). In the inset, the
  relaxation time $\tau$ obtained by interpolation of the correlation
  function as $e^{-\big({t-s\over\tau}\big)^\kappa}$ is plotted versus the inverse
  temperature $\beta$. The straight line is the interpolated
  behavior $\tau\sim e^{2.79(2)\beta}$.}
\label{Fig4}
\end{figure}
\end{center}

Finally, the stretched exponential scenario is tested on figure
\ref{Fig4}. The expected power-law behavior in the long-time regime
(according to equation \ref{eqStr}) is observed only for a narrow interval of
times, much narrower than in the two previous scenarii. We have nevertheless
interpolated the data with equation (\ref{eqStr}) to extract the relaxation
time. The result is plotted in the inset of figure \ref{Fig4}. An
exponential growth is observed but with a quite different factor:
$\tau\sim e^{2.79(2)\beta}$.

\section{Relaxation of the FFIM}
\begin{center}
\begin{figure}
\centerline{\includegraphics[width=12cm]{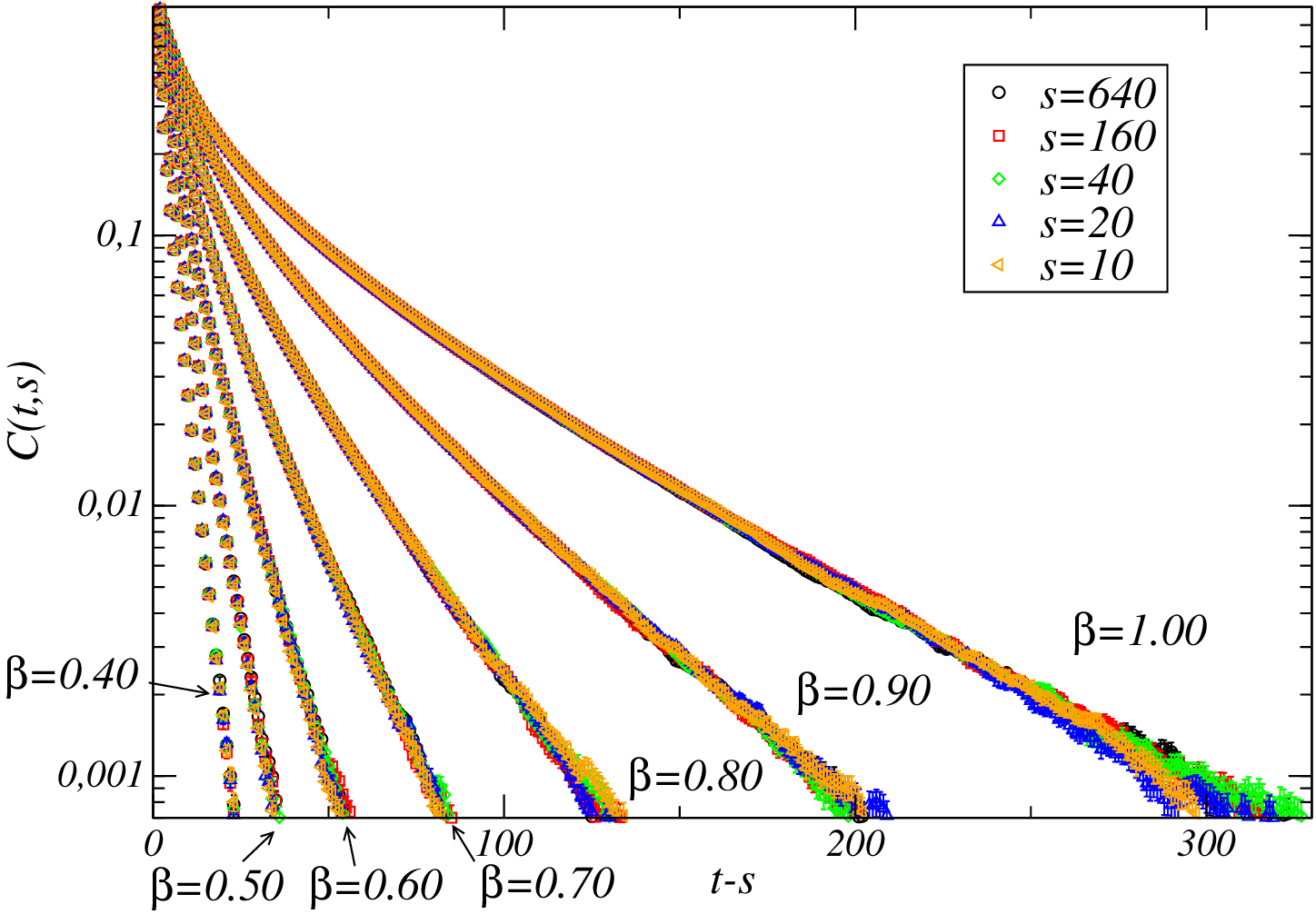}}
\caption{Relaxation of the spin-spin correlation function $C(t,s)$
of the FFIM versus $t-s$ at six different inverse temperatures
$\beta=0.50$, $0.60$, $0.70$, $0.80$, $0.90$ and $1.00$.
The different colors correspond to different times $s$ as indicated in
the legend.}
\label{Fig5}
\end{figure}
\end{center}

The FFIM has been studied in the zig-zag bond configuration for a
$192\times 192$ square lattice. We restricted ourselves to inverse temperatures
smaller or equal to $\beta=1$. As a consequence, smaller times and fewer
histories were necessary. Monte Carlo data have been averaged over $1,000$
independent histories. On figure \ref{Fig5}, spin-spin autocorrelation
functions $C(t,s)$ are presented for different temperatures. The collapse
of the different curves indicates that thermalization is achieved very
rapidly. Like for the AFIM, we have compared the three scenarii:
exponential decay of the spin-spin autocorrelation functions
(\ref{eq2}), exponential decay with logarithmic corrections
(\ref{eq4}), and stretched exponential (\ref{eqStr}). 

\begin{center}
\begin{figure}
\centerline{\includegraphics[width=12cm]{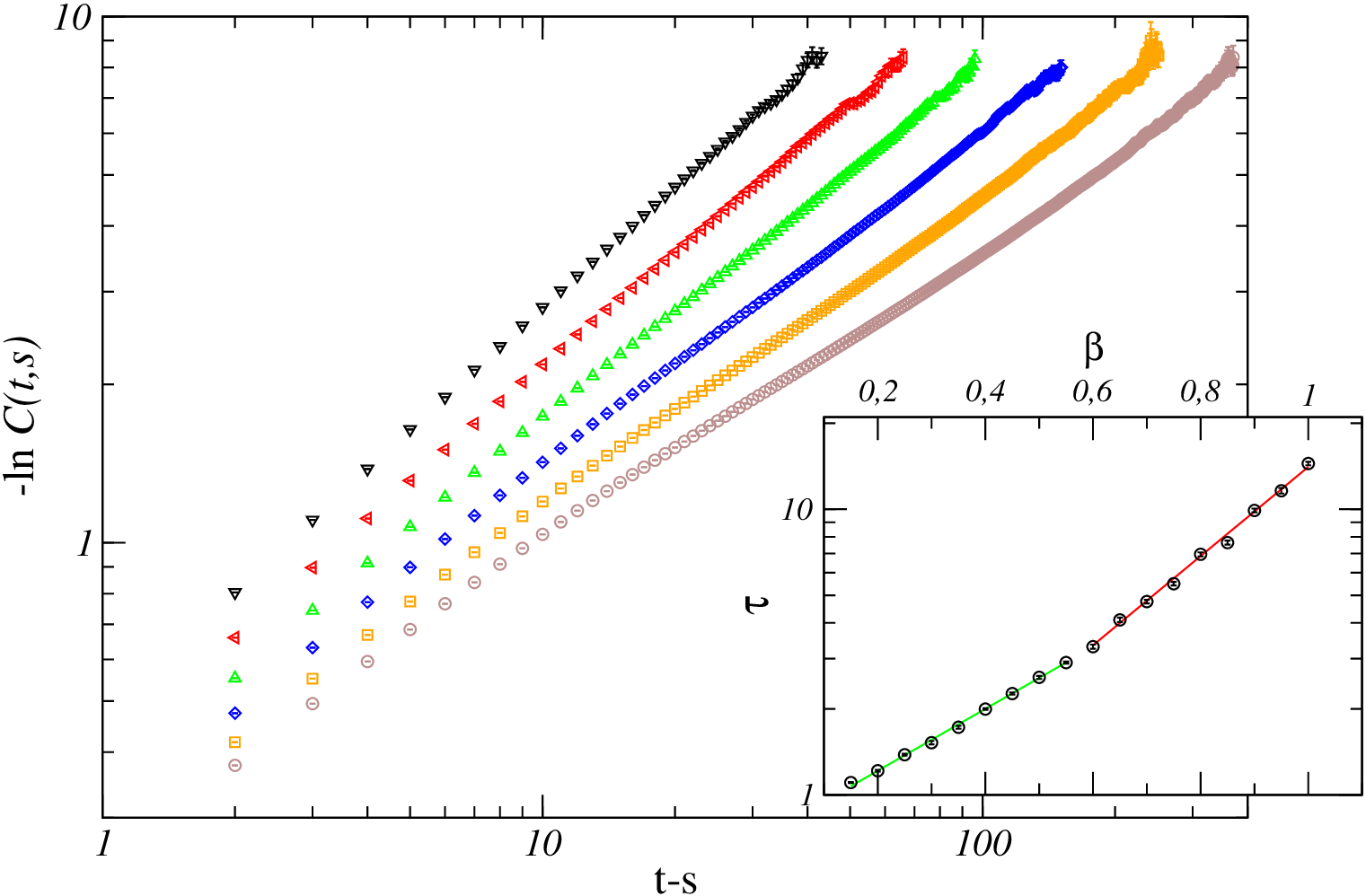}}
\caption{Logarithm $-\ln C(t,s)$ of the spin-spin correlation of the
  FFIM at different inverse temperatures $\beta=0.50$, $0.60$, $0.70$,
  $0.80$, $0.90$, and $1.00$ (from top to bottom).
  Only the time $s=160$ is presented.
  In the inset, the relaxation time $\tau$ obtained by interpolation of
  the correlation function as $e^{-\big({t-s\over\tau}\big)^\kappa}$ is plotted versus
  the inverse temperature $\beta$. The two straight lines are the interpolated
  behaviors $\tau\sim e^{3.60(5)\beta}$ for $\beta\ge 0.6$ and
  $\tau\sim e^{2.50(2)\beta}$ for $\beta<0.6$.
}
\label{Fig6}
\end{figure}
\end{center}

\begin{center}
\begin{figure}
\centerline{\includegraphics[width=12cm]{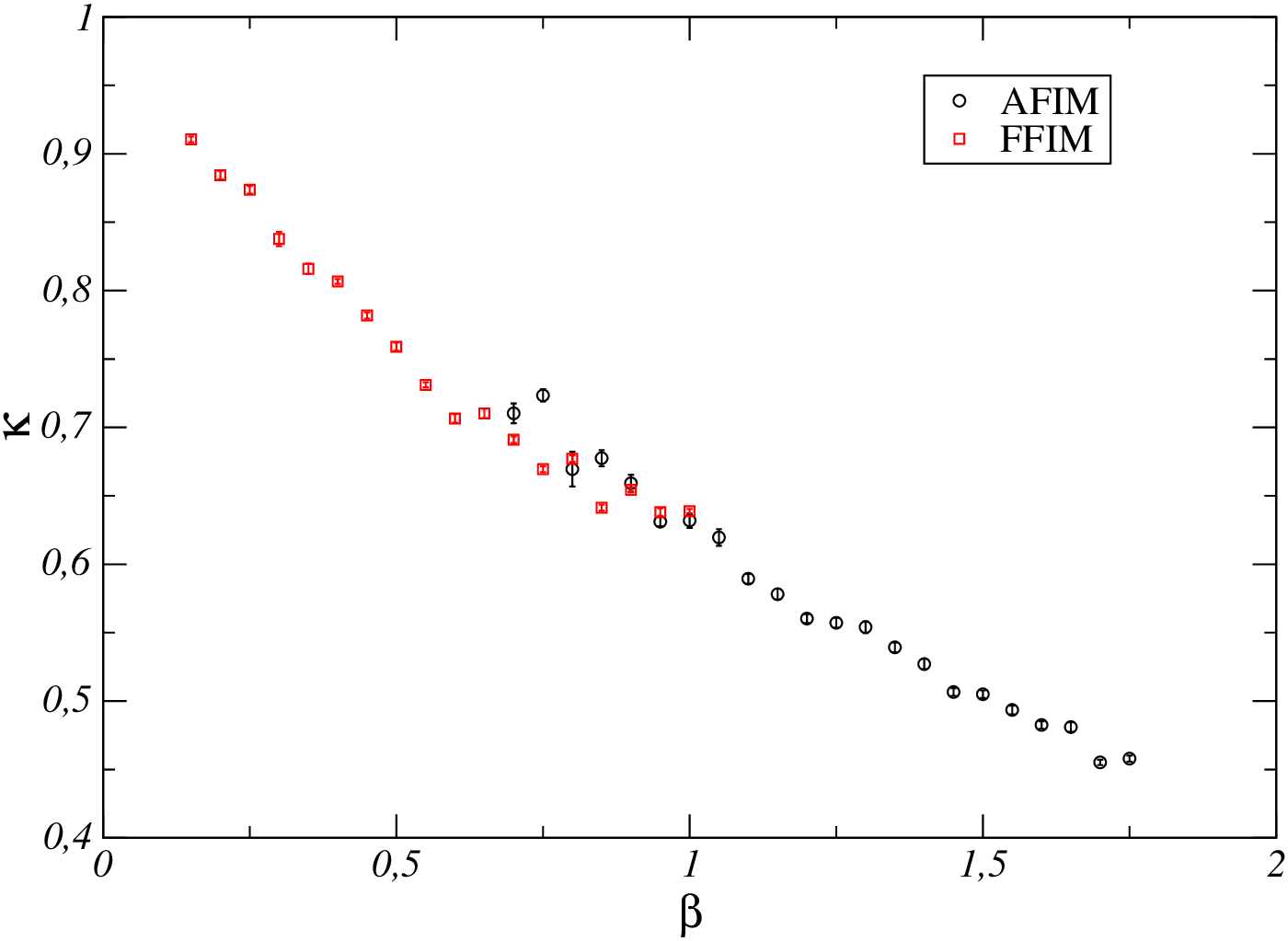}}
\caption{Exponent $\kappa$ of the stretched exponential versus
  the inverse temperature $\beta$ for both the AFIM (black)
  and the FFIM (red).}
\label{Fig6b}
\end{figure}
\end{center}

First, we present the test of the stretched-exponential scenario for the
FFIM (figure~\ref{Fig6}). The logarithm $-\ln C(t,s)$ displays a
behavior similar to the AFIM. At first sight, the power-law
behavior that is expected according to equation (\ref{eqStr})
seems to be observed over a larger range of times $t-s$ than for
the AFIM but one should keep in mind that the inverse temperatures
are different on figures \ref{Fig4} and \ref{Fig6}. 
We have estimated the exponent $\kappa$ of the stretched exponential.
Our data confirm the general trend observed in references
\cite{fierro_percolation_1999,franzese_precursor_1999}. The exponent
$\kappa$ indeed strongly depends on the temperature. However, we do
not recover a purely exponential decay above the percolation
temperature of the Fortuin-Kasteleyn clusters. We do not observe two
regimes ($\kappa=1$ for $T>T_p$ and $\kappa$ decreasing below $T_p$)
as previously found but instead, an exponent slowly approaching a
value $\kappa=1$ (see figure~\ref{Fig6b}). The discrepancy may be
explained by the larger lattice size used in this study ($L=192$
instead of $L=64$) and by the difficulty to identify a sufficiently
large power-law regime, especially at low temperature. As shown in the
inset of figure~\ref{Fig6}, the relaxation time $\tau$ does grow exponentially
over the whole range of inverse temperatures $\beta$ considered,
in contradistinction to the theoretical prediction (\ref{eq5}).
However, our data are compatible with two distinct regimes of exponential
growth with a prefactor estimated to be $3.60(5)$ for $\beta\ge 0.6$
and $2.50(2)$ for $\beta<0.6$ (note that $\beta_t\simeq 0.59$).

\begin{center}
\begin{figure}
\centerline{\includegraphics[width=12cm]{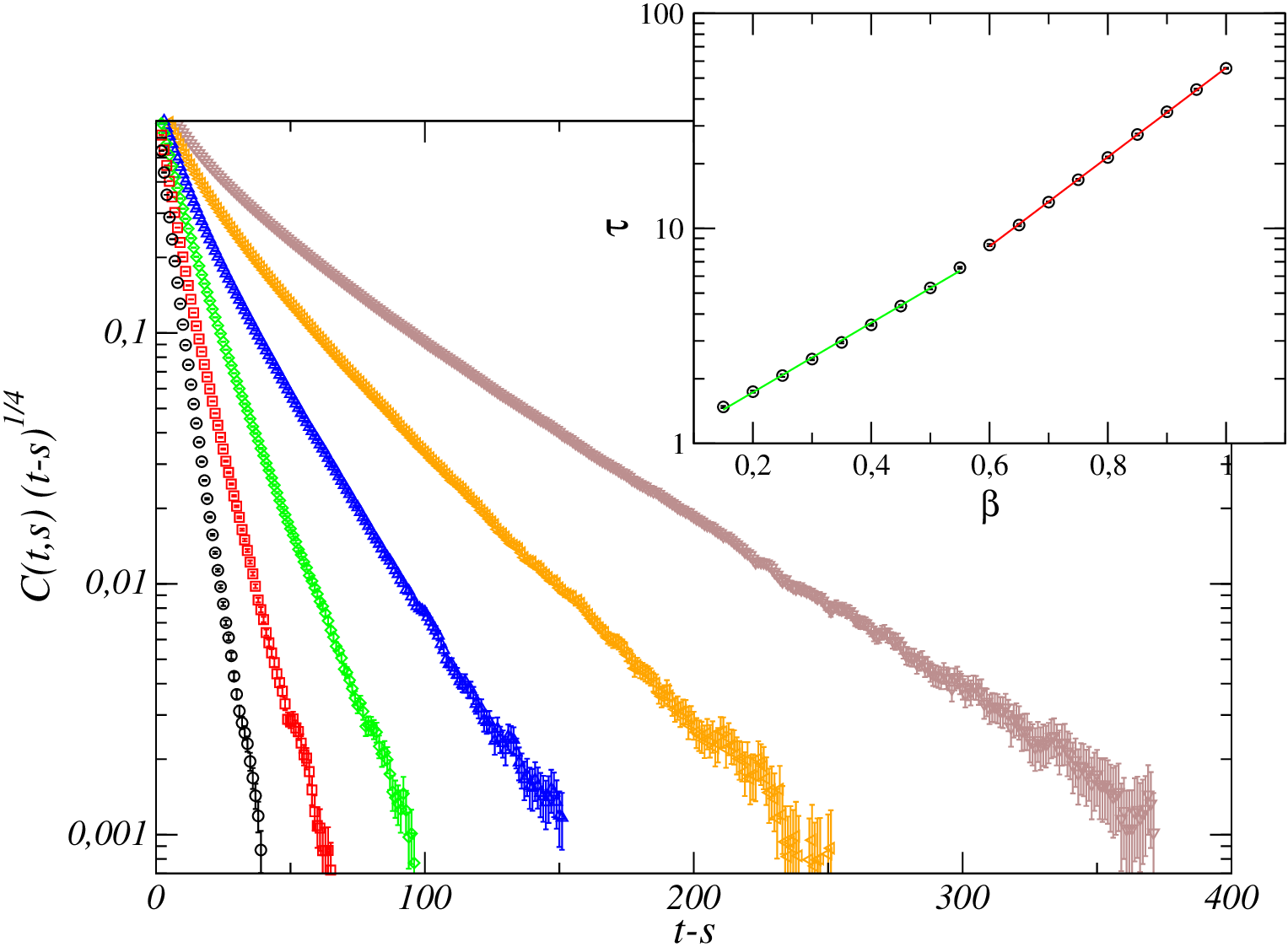}}
\caption{Scaling function $C(t,s)(t-s)^{1/4}$ at time $s=160$ versus
  $t-s$ for the FFIM at different inverse temperatures $\beta=0.50$,
  $0.60$, $0.70$, $0.80$, $0.90$, and $1.00$ (from left to right). In the
  inset, the relaxation time $\tau$ obtained by interpolation of the scaling
  function as $e^{-(t-s)/\tau}$ is plotted versus the inverse
  temperature $\beta$. The two straight lines are the interpolated
  behaviors $\tau\sim e^{4.78(2)\beta}$ for $\beta\ge 0.6$ and
  $\tau\sim e^{3.72(5)\beta}$ for $\beta<0.6$.}
\label{Fig7}
\end{figure}
\end{center}

On figure~\ref{Fig7}, the second scenario is tested. As expected, the
scaling function $C(t,s)(t-s)^{1/4}$ displays an exponential decay
with $t-s$ over a large range of times $t-s$ for different
temperatures. However, the relaxation time does not grow exponentially
over the whole range of temperatures considered (inset of figure
\ref{Fig7}). Again, two regimes can be distinguished: the prefactor in
the exponential (\ref{eq5}) is estimated to be $4.78(2)$ for $\beta\ge 0.6$
and $3.72(5)$ for $\beta<0.6$. These values should be compared with the
theoretical prediction $4$.
\smallskip

\begin{center}
\begin{figure}
\centerline{\includegraphics[width=12cm]{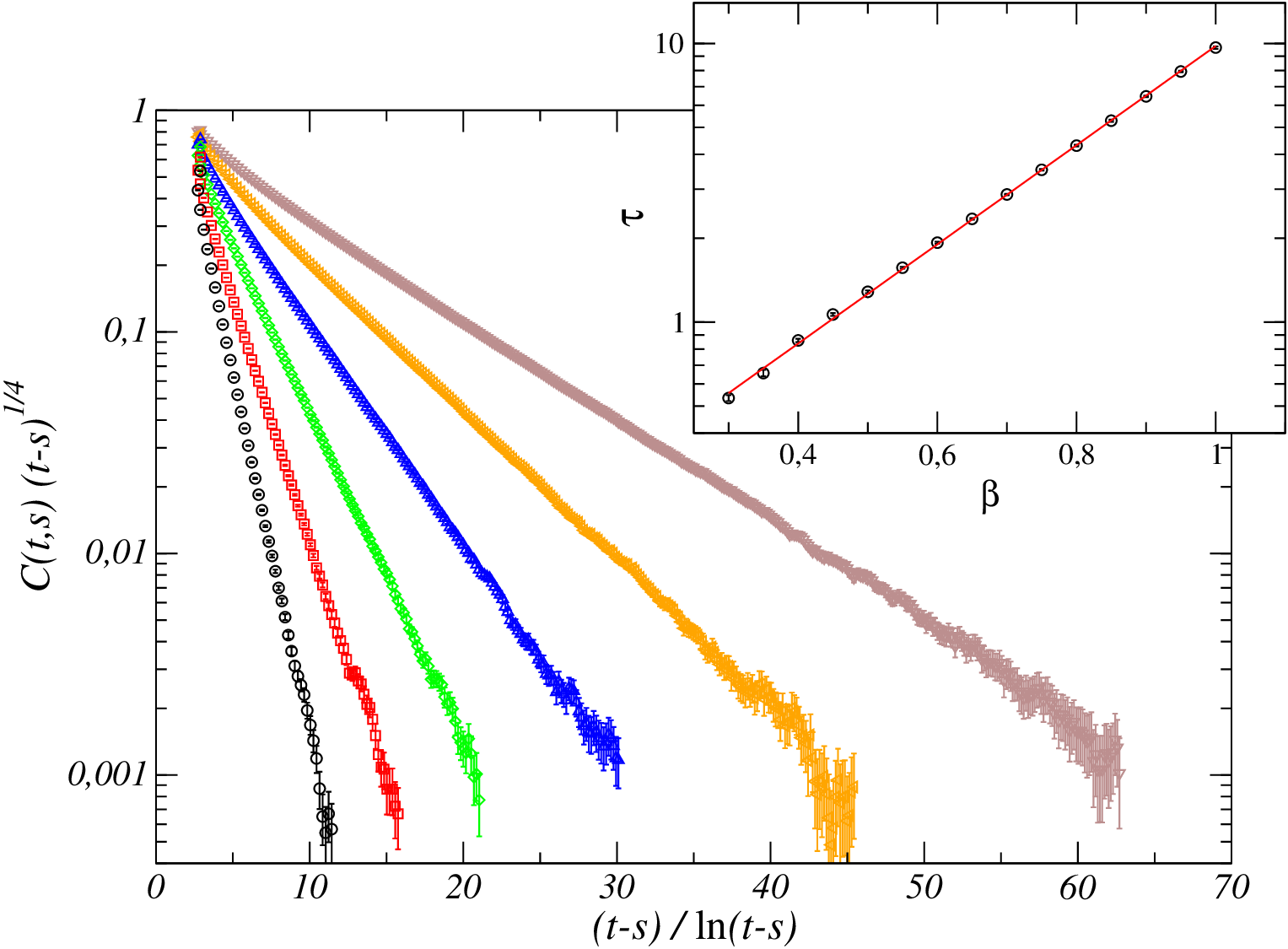}}
\caption{Scaling function $C(t,s)(t-s)^{1/4}$ at time $s=1600$ versus
  $(t-s)/\ln(t-s)$ for the AFIM at different inverse temperatures
  $\beta=0.50$, $0.60$, $0.70$, $0.80$, $0.90$, and $1.00$
  (from bottom to top). In the inset, the
  relaxation time $\tau$ obtained by interpolation of the scaling
  function as $e^{-(t-s)/\tau\ln(t-s)}$ is plotted versus the inverse
  temperature $\beta$.  The straight line is the interpolated
  behavior $\tau\sim e^{4.097(26)\beta}$.}
\label{Fig8}
\end{figure}
\end{center}

On figure~\ref{Fig8}, the third scenario involving logarithmic corrections
is tested. The scaling function $C(t,s)(t-s)^{1/4}$ decay exponentially
with $(t-s)/\ln(t-s)$ over a large range of times $t-s$ for different
temperatures. The relaxation time $\tau$ now displays an exponential growth
over the whole range of temperatures considered and our estimate
$4.097(26)$ of the prefactor in the argument of the exponential is much closer
to the theoretical prediction $4$ according to equation (\ref{eq5}) than
without logarithmic corrections.

\begin{table}[!ht]
\begin{center}
\begin{tabular}{@{}*{4}{l}}
\hline
Model & Stretched exponential & Exponential & Log. corrections \\
\hline
AFIM & $2.79(2)$ & $4.31(2)$ & $3.92(2)$ \\
FFIM & $3.60(5)$ ($\beta\ge 0.6$) & $4.78(2)$ ($\beta\ge 0.6$) & $4.097(26)$ \\
\hline
\end{tabular}
\end{center}
\caption{Prefactor in the argument of the exponential growth
(\ref{eq5}) of the relaxation time for the two models AFIM and FFIM
in the different studied scenarii. The theoretical prediction is $4$.}
\label{Table1}
\end{table}

\section{Conclusions}
We have analyzed the decay of the equilibrium two-time spin-spin
correlation functions in the paramagnetic phase against three
scenarii: exponential decay of the spin-spin autocorrelation functions
(\ref{eq2}), exponential decay with logarithmic corrections
(\ref{eq4}), and stretched exponential (\ref{eqStr}). To distinguish
between them, we tested the thermal behavior (\ref{eq5}) of the
relaxation time in the different scenarii. Our results are summarized in
table \ref{Table1}. The best agreement with the theoretical prediction
$4$ is obtained with an exponential decay with logarithmic corrections
(\ref{eq4}) for both the AFIM and the FFIM. Error bars are unfortunately
smaller than the deviation from the theoretical prediction. A great care
has been taken in the computation of error bars (from the data production
to the analysis) so we can only invoke finite-size effects or other
systematic deviations. The deviation for the two other scenarii being much
larger, we do not expect them to get closer to $4$ than the
exponential decay with logarithmic corrections.
These results bring further evidences of the existence of topological
defects in the paramagnetic phase of the AFIM and the FFIM.
\section*{References}
\bibliography{FSIM_HTv2}
\end{document}